\documentclass[aps,pra,twocolumn,superscriptaddress,10pt]{revtex4}
\usepackage{graphicx}
\usepackage{subfigure}
\usepackage{amsmath}
\usepackage{amsfonts}
\usepackage{amssymb}
\usepackage{amsthm}
\usepackage{times}
\usepackage[colorlinks,citecolor=blue,linkcolor=blue]{hyperref}
\usepackage{color}
\usepackage{setspace} 
\usepackage{verbatim}

\begin{document}
\title{Hybrid quantum-classical simulation of quantum speed limits in open quantum systems}

\newcommand{\UA}{\affiliation{Department of Chemistry, University of Alberta, Edmonton, Alberta, T6G 2G2, Canada }}

\newcommand{\UT}{\affiliation{Department of Chemistry and Centre for Quantum Information and Quantum Control, University of Toronto, 80 Saint George St., Toronto, Ontario, M5S 3H6, Canada}}

\author{Junjie Liu}
\UA
\author{Dvira Segal}
\UT
\author{Gabriel Hanna}
\email{gabriel.hanna@ualberta.ca}
\UA

\begin{abstract}
The quantum speed limit (QSL) provides a fundamental upper bound on the speed of quantum evolution, but its evaluation in generic open quantum systems still presents a formidable computational challenge. Herein, we introduce a hybrid quantum-classical method for computing QSL times in multi-level open quantum systems. The method is based on a mixed Wigner-Heisenberg representation of the composite quantum dynamics, in which the open subsystem of interest is treated quantum mechanically and the bath is treated in a classical-like fashion. By solving a set of coupled first-order deterministic differential equations for the quantum and classical degrees of freedom, one can compute the QSL time.  To demonstrate the utility of the method, we study the unbiased spin-boson model and provide a detailed analysis of the effect of the subsystem-bath coupling strength and bath temperature on the QSL time. In particular, we find a turnover of the QSL time in the strong coupling regime, which is indicative of a speed-up in the quantum evolution. We also apply the method to the Fenna-Matthews-Olson complex model and identify a potential connection between the QSL time and the efficiency of the excitation energy transfer at different temperatures.
\end{abstract}

\date{\today}

\maketitle

\section{Introduction}
The quantum speed limit (QSL) \cite{Mandelstam.45.JP,Margolus.98.PD,Deffner.17.JPA} sets the maximum speed (or equivalently the minimum time) at which a quantum system can evolve from an initial state to a target state. 
For closed systems and orthogonal states, the QSL time is given by the relation $\tau_{QSL}=\max\{\pi\hbar/(2\Delta E),\pi\hbar/(2E)\}$, with the two bounds usually referred to as Mandelstam-Tamm (MT) \cite{Mandelstam.45.JP} and Margolus-Levitin (ML) \cite{Margolus.98.PD} types, respectively. The MT bound depends on the variance of the energy of the initial state, $\Delta E^2$, while the ML bound depends on the mean energy with respect to the ground state, $E\equiv\langle G|\hat{H}|G\rangle$ (with $|G\rangle$ the ground state and $\hat{H}$ the Hamiltonian). Generalizations of the MT and ML bounds to nonorthogonal states and mixed initial states can be found in Refs. \cite{Pfeifer.93.PRL,Zwierz.12.PRA,Barnes.13.PRA,Poggi.13.EPL,Deffner.13.JPA,Russell.14.PRA,Campaioli.18.PRL}. The QSL has proven to be a useful concept in the fields of quantum computation \cite{Lloyd.00.N,Santos.15.SR}, quantum control \cite{Caneva.09.PRL,Hegerfeldt.13.PRL,Mukherjee.13.PRA,Hegerfeldt.14.PRA,Deffner.14.JPB,Campbell.17.PRL}, quantum metrology \cite{Giovannetti.11.NP,Chin.12.PRL,Tsang.13.NJP}, and quantum thermodynamics \cite{Deffner.10.PRL,Abah.17.EPL,Campaioli.17.PRL}.  Interestingly, recent studies on closed systems have shown that the QSL can have a classical counterpart \cite{Shanahan.18.PRL,Okuyama.18.PRL}.

In recent years, there has been a great deal of interest in the QSLs of open quantum systems (OQSs) \cite{Taddei.13.PRL,Del.13.PRL,Deffner.13.PRL,Xu.14.PRA,Xu.14.CPL,Liu.15.PRA,Sun.15.SR,Hou.15.JPA,Cimmarusti.15.PRL,Marvian.15.PRL,Marvian.16.PRA,Mirkin.16.PRA,Mondal.16.PLA,Pires.16.PRX,Ektesabi.17.PRA,Campaioli.18.A,Funo.18.A,Wu.18.PRA}.  For an OQS whose reduced dynamics is governed by a generic time-convolutionless master equation for the reduced density matrix, $\frac{d}{dt}\hat{\rho}_t=\hat{\mathcal{L}}_t\hat{\rho}_t$ (where $\hat{\mathcal{L}}_t$ is the time-dependent evolution superoperator) \cite{Breuer.07.NULL}, the QSL time can be expressed as \cite{Deffner.13.PRL,Deffner.17.NJP}
\begin{equation}\label{eq:qsl_definition}
\tau_{QSL}~=~\mathrm{max}\{\tau_1,\tau_2,\tau_{\infty}\},
\end{equation}
where $\tau_p=\sin^2[\mathcal{B}(\hat{\rho}_0,\hat{\rho}_{\tau})]/E_{\tau}^p$.  The numerator of $\tau_p$ contains the Bures angle, $\mathcal{B}(\hat{\rho}_0,\hat{\rho}_{\tau})=\arccos(\sqrt{\mathcal{F}(\hat{\rho}_0,\hat{\rho}_{\tau})})$, between an initial state $\hat{\rho}_0$ and a final state $\hat{\rho}_{\tau}$ at time $\tau$ \cite{Bures.69.TAMS}, where $\mathcal{F}=\left[\mathrm{Tr}_S\sqrt{\sqrt{\hat{\rho}_0}\hat{\rho}_{\tau}\sqrt{\hat{\rho}_0}}\right]^2$ is the quantum fidelity \cite{Jozsa.94.JMO}. The denominator $E_{\tau}^p=\frac{1}{\tau}\int_0^{\tau}dt||\hat{\mathcal{L}}_t\hat{\rho}_t||_p$ is the time average of $||\hat{\mathcal{L}}_t\hat{\rho}_t||_p$ over an actual evolution time interval $\tau$, where $||\hat{A}||_p=\left(\sum_k\alpha_k^p\right)^{1/p}$ is the Schatten-$p$-norm of $\hat{A}$ with $\alpha_k$ the $k^{th}$ singular value of $\hat{A}$, i.e., the $k^{th}$ eigenvalue of the Hermitian operator $\sqrt{\hat{A}^{\dagger}\hat{A}}$ \cite{Bhatia.97.NULL}.  Thus, Eq.~(\ref{eq:qsl_definition}) involves Schatten-$p$-norms with $p=1,2$, and $\infty$ (referred to as the familiar trace, Hilbert-Schmidt, and operator norms, respectively). It should be noted that $\tau_{QSL}$ is not a physical time, but rather an intrinsic characteristic timescale associated with a system's dynamics that satisfies the bound $\tau_{QSL}\le\tau$ \cite{Deffner.17.JPA}. 

Because $\tau_{QSL}$ explicitly depends on the physical time $\tau$, it can be evaluated in two possible ways. The first way, which involves fixing the evolution time $\tau$, has been used to identify memory effects on the speed of quantum evolution \cite{Deffner.13.PRL,Xu.14.PRA}. In contrast, the second way involves varying the evolution time $\tau$ and studying $\tau_{QSL}$ as a function of $\tau$. This approach has been employed in the study of entanglement-assisted speed-up of quantum evolution \cite{Giovannetti.03.PRA,Batle.05.PRA,Borras.06.PRA,Frowis.12.PRA}.  In both cases, the smaller the value of $\tau_{QSL}$ is, the faster the quantum evolution can be realized. Therefore, $\tau_{QSL}$ could potentially be used as a performance metric in the design of quantum-based technologies.    

It was previously shown that the operator norm $||\hat{\mathcal{L}}_t\hat{\rho}_t||_{\infty}$ yields the maximum value of $\tau_{QSL}$ in Eq.~(\ref{eq:qsl_definition}) \cite{Deffner.13.PRL}. However, computing the operator norm is far from being a trivial task, as extracting the largest singular value from a time-dependent reduced density matrix is computationally expensive, if at all feasible. Thus far, evaluations of $\tau_{QSL}$ in OQSs have been limited to either exactly solvable models \cite{Del.13.PRL,Deffner.13.PRL,Xu.14.PRA,Xu.14.CPL,Liu.15.PRA,Cheng.18.A,Deffner.17.NJP}, or have involved approximated forms of time-local master equations with limited applicabilities \cite{Mukherjee.13.PRA,Cimmarusti.15.PRL,Funo.18.A}. Recently, an alternative expression for $\tau_{QSL}$ in Wigner phase space was proposed in Ref.~\cite{Deffner.17.NJP}, which circumvents the task of determining the singular values of a high-dimensional operator. Nevertheless, determining the Wigner function for an arbitrary OQS is very challenging. Intriguing aspects of the QSL time have been revealed in previous studies on small and exactly solvable systems \cite{Del.13.PRL,Deffner.13.PRL,Xu.14.PRA,Xu.14.CPL,Liu.15.PRA,Cheng.18.A,Deffner.17.NJP}.  However, moving forward, it is desirable to develop computationally efficient methods for evaluating $\tau_{QSL}$ in arbitrary OQSs, especially considering the fact that knowledge of the minimal duration of a process is of fundamental importance to virtually all areas of quantum physics.

To address this challenge, we propose herein to compute $\tau_{QSL}$ for multi-level OQSs using a hybrid quantum-classical method \cite{Liu.18.JPCL}, which relies on the mixed Wigner-Heisenberg representation \cite{Aleksandrov.81.ZNA,Gerasimenko.82.TMP,Zhang.88.JPP,Kapral.99.JCP} of the composite quantum dynamics. In contrast to Refs.~\cite{Deffner.17.NJP,Shanahan.18.PRL}, we apply the Wigner transform only to the bath degrees of freedom (yielding a classical-like description of the bath) and retain the operator character of the subsystem degrees of freedom. In doing so, one can treat OQSs with large numbers of bath degrees of freedom and complex bath models, i.e., situations where exact methodologies become computationally intractable. 

In this work, without loss of generality, we focus on $\tau_2$ in Eq.~(\ref{eq:qsl_definition}) because the Hilbert-Schmidt norm $||\hat{\mathcal{L}}_t\hat{\rho}_t||_2$ is mathematically less involved than the other measures, and can be expressed as $\sqrt{\mathrm{Tr}_S\left(\frac{d}{dt}\hat{\rho}_t\right)^2}$ \cite{Deffner.13.PRL}. We note that $\tau_2$ can also be obtained as a limit of an improved bound for the QSL time, which relies on an alternative definition of the quantum fidelity between states \cite{Ektesabi.17.PRA}. The time-dependence of $\frac{d}{dt}\hat{\rho}_t$ can be readily simulated using our hybrid quantum-classical method without calculating singular values and without explicitly constructing the superoperator $\hat{\mathcal{L}}_t$. Although $\tau_2$ does not constitute the ``tightest" bound (i.e., the duration of the process always exceeds $\tau_2$), it can still yield the same qualitative information as the tightest bound $\tau_{\infty}$.  In some cases, $\tau_2$ may not even be much smaller than $\tau_{\infty}$, e.g., $\tau_2$ is only a factor of $\sqrt{2}$ smaller than $\tau_{\infty}$ for a two-level atom in a photonic crystal cavity \cite{Xu.14.PRA}.

We apply the hybrid quantum-classical method to two prototype models whose quantum dynamics have been extensively studied over the years: the spin-boson model (SBM) \cite{Leggett.87.RMP,Weiss.12.NULL} and the Fenna-Matthews-Olson (FMO) complex model \cite{Fenna.75.N,Adolphs.06.BJ,Ishizaki.09.PNAS}. For the SBM, we focus on the influence of the bath temperature and subsystem-bath coupling strength on the behaviour of $\tau_{QSL}$. 
We also compare our results with those obtained from the non-Markovian Bloch-Redfield equation (NM-BRE) and its Markovian version (M-BRE) \cite{Aslangul.86.JP,Thoss.01.JCP,Boudjada.14.JPCA}. 
For the FMO complex model, we explore the relationship between the QSL time and the efficiency of the excitation energy transfer at a physiological temperature (300 K) and a cryogenic temperature (77 K). 

The paper is organized as follows: In section \ref{sec:1} we present the working expressions for the QSL time and the hybrid quantum-classical method, and then illustrate how to compute the QSL time using this method. In sections \ref{sec:2} and \ref{sec:3}, we apply the hybrid quantum-classical approach to the SBM and FMO complex model. Here, the detailed simulation results and discussions are provided. We summarize our findings in section \ref{sec:4}. The hybrid quantum-classical equations of motion for the SBM and FMO complex model are provided in the Appendix.

\section{Working Expressions and Methodology}
\label{sec:1}

\subsection{Quantum speed limit time expression}

As mentioned in the introduction, we will use $\tau_2$ in Eq.~(\ref{eq:qsl_definition}) as a measure of the QSL time $\tau_{QSL}$. 
Noting the relation between the Bures angle and quantum fidelity, we therefore have
\begin{equation}\label{eq:tqsl}
\tau_{QSL}=\frac{1-\mathcal{F}(\hat{\rho}_0,\hat{\rho}_{\tau})}{\frac{1}{\tau}\int_0^{\tau}dt \sqrt{\mathrm{Tr}_S\left(\frac{d}{dt}\hat{\rho}_t\right)^2}}.
\end{equation}
It should be noted that although the above expression is derived by making use of the time-local master equation, it can be applied to arbitrary OQSs as it only depends on the reduced density matrix and its time derivative. Therefore, one can use any method, approximate or exact, that yields time-dependent information to evaluate $\tau_{QSL}$ in Eq.~(\ref{eq:tqsl}).

In this work, we take the initial condition of the multi-level subsystem to be a pure state, 
$\hat{\rho}_0=|n\rangle\langle n|$, where $\{|n\rangle\}$ spans the subsystem Hilbert space.  In this case, the quantum fidelity reduces to \cite{Jozsa.94.JMO}
\begin{equation}\label{eq:ff_definition}
\mathcal{F}(\hat{\rho}_0,\hat{\rho}_{\tau})~=~\langle \mathcal{\hat{P}}_{nn}(\tau)\rangle,
\end{equation}
where $\mathcal{\hat{P}}_{nn}=|n\rangle\langle n|$ is a subsystem projection operator.  
We further obtain that
\begin{eqnarray}\label{eq:r2}
\mathrm{Tr}_S\left(\frac{d}{dt}\hat{\rho}_t\right)^2 &=& \sum_{n=1}^L\sum_{m=1}^L\left|\frac{d}{dt}\rho_t^{mn}\right|^2\nonumber\\
&=&  \sum_{n=1}^L\sum_{m=1}^L\left|\frac{d}{dt}\langle \mathcal{\hat{P}}_{nm}(t)\rangle\right|^2,
\end{eqnarray}
where $L$ is the number of subsystem levels, $\rho_t^{mn}=\langle m|\hat{\rho}_t|n\rangle$, $|\cdot|$ takes the norm of its argument, and we have used the fact that 
\begin{equation} \label{eq:rhor}
\rho_t^{mn}=\mathrm{Tr}_S\left[\mathcal{\hat{P}}_{nm}\hat{\rho}_t\right]=\langle \mathcal{\hat{P}}_{nm}(t)\rangle.
\end{equation}
From Eqs.~(\ref{eq:ff_definition}) and (\ref{eq:r2}), we see that the QSL time in Eq.~(\ref{eq:tqsl}) is fully determined by ensemble averages of time-dependent projection operators and their time derivatives. Below, we show how to compute these ensemble averages using a hybrid quantum-classical dynamics method.

\subsection{The hybrid quantum-classical formalism and the DECIDE implementation}
To compute the QSL time in an OQS, one needs a methodology that can accurately capture the reduced dynamics of a quantum subsystem. Such time evolution methods range from numerically exact techniques, such as the quasiadiabatic propagator path integral (QUAPI) \cite{Makri.95.JCP,Makri.95.JCPa} approach, to weak-system-bath perturbative methods that
are derived systematically from the Nakajima Zwanzig equation \cite{Breuer.07.NULL}.
While path integral-based tools offer an exact numerical solution, they are limited
to treating small systems due to their computational costs. As well, converging the dynamics in ``difficult" parameter regimes (i.e., strong subsystem-bath coupling, low temperature) becomes increasingly challenging. On the other end, the perturbative Redfield equation is easy to implement and employ, but given its perturbative nature it can only accurately capture weak subsystem-bath coupling effects. 

Hybrid quantum-classical methods \cite{Tully.90.JCP,Billing.93.JCP,Prezhdo.97.PRA,Martens.97.JCP,Donoso.98.JPCA,Tully.98.FD,Kapral.99.JCP,Donoso.00.JCP,Wan.00.JCP,Horenko.02.JCP,Wan.02.JCP,Mackernan:2002,Horenko.04.JCP,Roman.07.JPCA,kim08b,mackernan08,Bai.14.JPCA,Kim.14.JCP,Kim.14.JCPa,Wang.15.JPCL,Martens.16.JPCL,Wang.16.JPCL,Agostini.16.JCTC,Subotnik.16.ARPC,Kapral.15.JP}, which treat the subsystem of interest quantum mechanically and the bath in a classical-like fashion, are viable alternatives to fully quantum mechanical ones. These methods are particularly useful for modelling quantum dynamical processes occurring in condensed phases. 
To arrive at a hybrid quantum-classical description of the dynamics, one can first perform a partial Wigner transform \cite{Wigner.32.PR} over the bath degrees of freedom of the quantum Liouville equation, which introduces a phase space description of the bath variables while retaining the operator character of the subsystem degrees of freedom, and then make physically motivated approximations.  For example, by linearizing the resulting equation in $\hbar$, one can obtain the quantum-classical Liouville equation \cite{Aleksandrov.81.ZNA,Gerasimenko.82.TMP,Zhang.88.JPP,Kapral.99.JCP}.  In this work, we adopt a recently developed hybrid quantum-classical approach \cite{Liu.18.JPCL} that solves the quantum-classical Liouville equation in an efficient manner; namely, solving a system of coupled first-order differential equations (FODEs) for the coordinates of the subsystem and bath. We now describe the approach and its implementation.

Let us consider a generic OQS described by the following Hamiltonian
\begin{equation}\label{eq:hh}
\hat{H}~=~\hat{H}_S(\boldsymbol{\hat{x}})+\hat{H}_I(\boldsymbol{\hat{x}},\boldsymbol{\hat{X}})+\hat{H}_{B}(\boldsymbol{\hat{X}}).
\end{equation}
Here, $\hat{H}_S$ is the subsystem Hamiltonian and $\boldsymbol{\hat{x}}$ collectively refers to the complete set of subsystem projection operators $\mathcal{\hat{P}}_{nm}$ (i.e., the generalized coordinates of the subsystem); $\hat{H}_B=\sum_{j=1}^{N}[\hat{P}_{j}^2/2+\omega_{j}^2\hat{R}_{j}^2/2]$ is the Hamiltonian of the bosonic heat bath (containing $N$ harmonic oscillators) characterized by a temperature $T$ (or inverse temperature $\beta\equiv1/T$), where $\hat{P}_{j}$, $\hat{R}_{j}$, and $\omega_{j}$ are the mass-weighted momentum, position, and frequency of $j$th oscillator, respectively, and $\boldsymbol{\hat{X}}=(\boldsymbol{\hat{R}},\boldsymbol{\hat{P}})$ with $\boldsymbol{\hat{R}}=(\hat{R}_{1},\hat{R}_{2},\ldots,\hat{R}_{N})$ and $\boldsymbol{\hat{P}}=(\hat{P}_{1},\hat{P}_{2},\ldots,\hat{P}_{N})$; and $\hat{H}_I$ is the subsystem-bath interaction Hamiltonian. The extension to multiple heat baths is straightforward, as will be seen.  In what follows, we set $\hbar=1$ and $k_B=1$.

Since the generalized coordinates $\boldsymbol{\hat{x}}$ can be used to construct the reduced density matrix through Eq.~(\ref{eq:rhor}), their time evolution can be used to determine the QSL time $\tau_{QSL}$. To simulate the time evolution $\boldsymbol{\hat{x}}$, we adopt the mixed Wigner-Heisenberg representation of the composite quantum dynamics \cite{Aleksandrov.81.ZNA,Gerasimenko.82.TMP,Zhang.88.JPP,Kapral.99.JCP}, which involves performing partial Wigner transforms over the bath coordinates of bath-dependent operators $\hat{A}(\boldsymbol{\hat{X}})$, i.e., 
\begin{equation}
\hat{A}_W(\boldsymbol{X})~=~\int\,d\boldsymbol{z}e^{i\boldsymbol{P}\cdot\boldsymbol{z}/\hbar}\left\langle \boldsymbol{R}-\frac{\boldsymbol{z}}{2}\left|\hat{A}\right|\boldsymbol{R}+\frac{\boldsymbol{z}}{2}\right\rangle,
\end{equation}
where $\boldsymbol{X}=(\boldsymbol{R},\boldsymbol{P})$ denotes the bath phase space variables and the subscript $W$ indicates that a partial Wigner transform has been performed. After this transformation, the composite dynamics is equivalently governed by the following Weyl-ordered, mixed Wigner-Heisenberg form of the Hamiltonian
\begin{equation}
\hat{H}_W(\boldsymbol{\hat{x}},\boldsymbol{X})~=~\hat{H}_S(\boldsymbol{\hat{x}})+\hat{H}_{I,W}(\boldsymbol{\hat{x}},\boldsymbol{X})+H_{B,W}(\boldsymbol{X}).
\end{equation}
Weyl-ordering involves replacing product terms such as $\boldsymbol{\hat{x}}\boldsymbol{X}$ in $\hat{H}_W$ with the expression $\frac{1}{2}(\boldsymbol{\hat{x}}\boldsymbol{X}+\boldsymbol{X}\boldsymbol{\hat{x}})$.  In this work, we simulate the dynamics of $\boldsymbol{\hat{x}}(t)$ and $\boldsymbol{X}(t)$ using the so-called DECIDE (\textbf{D}eterministic \textbf{E}volution of \textbf{C}oordinates with \textbf{I}nitial \textbf{D}ecoupled \textbf{E}quations) scheme recently proposed in Ref.~\cite{Liu.18.JPCL}, which provides an efficient, albeit approximate, way of solving the quantum-classical Liouville equation for $\hat{A}_W(\boldsymbol{X},t)$.  By assuming a factorized initial state for the composite system (namely, $\hat{\rho}_{tot}(0)=\hat{\rho}_0\otimes\hat{\rho}_B$ where $\hat{\rho}_{tot}$ and $\hat{\rho}_{B}=\frac{e^{-\beta\hat{H}_B}}{\mathrm{Tr}_B[e^{-\beta\hat{H}_B}]}$ are the density matrices of the total system and bath, respectively), it was shown that the quantum-classical Liouville equation for $\hat{A}_W(\boldsymbol{X},t)$ could be unfolded into the following approximate set of coupled equations of motion (EOMs) for the time-dependent coordinates $\boldsymbol{\hat{x}}(t)$ and $\boldsymbol{X}(t)$ (see Ref.~\cite{Liu.18.JPCL} for the details of the derivation and the approximations involved),
\begin{eqnarray}\label{eq:tt_final}
\frac{d}{dt}\boldsymbol{\hat{x}}(t) &=& i\left([\hat{H}_W,\boldsymbol{\hat{x}}]\right)(t),\nonumber\\
\frac{d}{dt}\boldsymbol{X}(t) &=& -\left(\{\hat{H}_W,\boldsymbol{X}\}\right)(t),
\end{eqnarray}
where $[\cdot,\cdot]$ and $\{\cdot,\cdot\}$ denote a commutator and Poisson bracket, respectively.
To treat the quantum operators and classical variables on an equal footing, we cast Eq.~(\ref{eq:tt_final}) in a convenient basis $\{|\alpha\rangle\}=(|\alpha_1\rangle,\ldots,|\alpha_L\rangle)$ that spans the Hilbert space of the $L$-dimensional quantum subsystem,  
\begin{eqnarray}\label{eq:eom_detail}
\frac{d}{dt}\boldsymbol{x}^{\alpha\alpha^{\prime}}(t) &=& i\langle\alpha|\left([\hat{H}_W, \boldsymbol{\hat{x}}]\right)(t)|\alpha^{\prime}\rangle,\nonumber\\
\frac{d}{dt}\boldsymbol{X}^{\alpha\alpha^{\prime}}(t) &=& -\langle\alpha|\left(\{\hat{H}_W, \boldsymbol{X}\}_a\right)(t)|\alpha^{\prime}\rangle,
\end{eqnarray}
where $D^{\alpha\alpha^{\prime}}\equiv\langle\alpha|D|\alpha^{\prime}\rangle$.
For bilinear subsystem-bath interactions, $i\langle\alpha|\left([\hat{H}_W, \boldsymbol{\hat{x}}]\right)(t)|\alpha^{\prime}\rangle$ becomes a functional of the matrix elements $\{\boldsymbol{x}^{\alpha\alpha^{\prime}}(t)\}$ and $\{(\boldsymbol{\hat{x}}(t)\boldsymbol{X}(t)+\boldsymbol{X}(t)\boldsymbol{\hat{x}}(t))^{\alpha\alpha^{\prime}}\}$, the latter arising from the bilinear interaction in the Weyl-ordered Hamiltonian $\hat{H}_W$.  On the other hand, $-\langle\alpha|\left(\{\hat{H}_W, \boldsymbol{X}\}_a\right)(t)|\alpha^{\prime}\rangle$ becomes a functional of the matrix elements $\{\boldsymbol{x}^{\alpha\alpha^{\prime}}(t)\}$ and $\{\boldsymbol{X}^{\alpha\alpha^{\prime}}(t)\}$. 
The notation $\{\boldsymbol{z}^{\alpha\alpha^{\prime}}\}$ denotes a particular set of matrix elements of $\boldsymbol{z}$ in the basis $\{|\alpha\rangle\}$, the contents of which depend on the model under investigation.  Also, $(\hat{x}_lX_k)^{\alpha\alpha^{\prime}}=\sum_{\beta}x_l^{\alpha\beta}X_k^{\beta\alpha^{\prime}}$. 
It should be noted that, due to the subsystem-bath interactions, the classical coordinates depend on the subsystem operators and $\boldsymbol{X}^{\alpha\alpha^{\prime}}(t)\neq \boldsymbol{X}(t)\delta_{\alpha\alpha^{\prime}}$ at finite times (where $\delta_{\alpha\alpha^{\prime}}$ is the Kronecker delta function). 
The superscript on $\boldsymbol{X}^{\alpha\alpha^{\prime}}(t)$ acts as a label to distinguish the various $\it{c}$-numbers and their corresponding EOMs at finite times.  

Equation (\ref{eq:eom_detail}) constitutes a set of coupled FODEs for the {\it c}-numbers ($\boldsymbol{x}^{\{\alpha\alpha^{\prime}\}}(t),\boldsymbol{X}^{\{\alpha\alpha^{\prime}\}}(t)$), where $\{\alpha\alpha^{\prime}\}$ denotes all the combinations of basis indices. The maximum number of coupled FODEs is $L^2(L^2-1+2N)$:
the subsystem is described by $L^2-1$ projection operators (because the identity operator is excluded), the $N$ harmonic oscillators are 
described by $N$ displacements and $N$ momenta, and each of the resulting $(L^2-1+2N)$ coordinates has $L^2$ equations.
However, one could reduce the number of FODEs if the subsystem has some symmetry. It should be noted that DECIDE is not limited to bilinear interactions, provided that the interaction can be decomposed into a finite number of terms involving matrix elements of coordinates. 

The domain of applicability of DECIDE warrants some comments. To arrive at Eq.~(\ref{eq:tt_final}), one must truncate the corresponding quantum Heisenberg equations for the coordinates (see Ref.~\cite{Liu.18.JPCL} for details).  In doing so, one neglects higher order terms that contain derivatives of the time-dependent coordinates with respect to the initial bath coordinates. As a result, one can underestimate the back-action from the bath onto the subsystem. If the subsystem dynamics is {\it highly} non-Markovian, these higher order terms are important and DECIDE will yield inaccurate results in the long-time limit. We note that strong memory effects can be induced by strong subsystem-bath couplings, slow baths with characteristic timescales longer than that of the subsystem, and very low temperatures. 
Therefore, DECIDE should be used with caution in these regimes.

\subsection{Evaluation of quantum speed limit time}
The ensemble averages of time-dependent projection operators and their time derivatives found in the QSL time expression in Eq.~(\ref{eq:tqsl}) have the following forms in the mixed Wigner-Heisenberg representation \cite{Sergi.03.TCA}
\begin{eqnarray}\label{eq:ea}
&&\frac{d}{dt}\langle \mathcal{\hat{P}}_{nm}(t)\rangle =\sum_{\alpha\alpha^{\prime}}\int\,d\boldsymbol{X}(0)\rho_{B,W}(\boldsymbol{X}(0))\frac{d}{dt}\mathcal{P}_{nm}^{\alpha\alpha^{\prime}}(t)\rho_0^{\alpha^{\prime}\alpha},\nonumber\\
&&\langle \mathcal{\hat{P}}_{nm}(t)\rangle = \sum_{\alpha\alpha^{\prime}}\int\,d\boldsymbol{X}(0)\rho_{B,W}(\boldsymbol{X}(0))\mathcal{P}_{nm}^{\alpha\alpha^{\prime}}(t)\rho_0^{\alpha^{\prime}\alpha}
\end{eqnarray}
where $\rho_{B,W}(0)$ is the partially Wigner-transformed thermal equilibrium distribution \cite{Imre.67.JMP}
\begin{eqnarray}\label{eq:rhoBw}
\rho_{B,W}(0) &=& \prod_{j=1}^N\frac{\tanh(\beta\omega_j/2)}{\pi}\exp\left[-\frac{2\tanh(\beta\omega_j/2)}{\omega_j}\right.\nonumber\\
&&\left.\times \left(\frac{P_j^2}{2}+\frac{\omega_j^2R_j^2}{2}\right)\right],
\end{eqnarray}
satisfying $\int d\boldsymbol{X}(0)\rho_{B,W}(\boldsymbol{X}(0))=1$.

Equation (\ref{eq:ea}) suggests a trajectory-based molecular dynamics (MD) approach to compute the QSL time $\tau_{QSL}$:  One generates a swarm of independent classical-like trajectories starting from different $\boldsymbol{X}(0)$ sampled from $\rho_{B,W}(0)$, and the same initial values of the matrix elements of the projection operators. Each trajectory of $\mathcal{P}_{nm}^{\alpha\alpha^{\prime}}(t)$ is obtained by integrating the $L^2(L^2-1+2N)$ coupled FODEs in Eq.~(\ref{eq:eom_detail}) using the standard fourth-order Runge-Kutta scheme \cite{Dormand.80.JCAM}. Inserting the time-dependent coordinates back into Eq.~(\ref{eq:eom_detail}) yields the corresponding time derivatives. Averaging the $\mathcal{P}_{nm}^{\alpha\alpha^{\prime}}(t)$ and their time derivatives over the ensemble of trajectories yields the required ensemble averages found in the QSL time.  Finally, Simpson's rule \cite{Suli.03.NULL} is used to perform the numerical integration in the denominator of Eq.~(\ref{eq:tqsl}). To illustrate our methodology, we study the SBM and FMO complex model in the following sections.

\section{The spin-boson model}\label{sec:2}

The behaviour of the QSL time has been explored in several exactly-solvable OQS models, including the damped Jaynes-Cummings model and the pure dephasing model \cite{Deffner.13.PRL,Cheng.18.A}.  Herein, using the DECIDE hybrid quantum-classical method, we study the behaviour of the QSL time in the spin-boson model \cite{Leggett.87.RMP}, which exhibits a rich dynamics 
but lacks a closed analytic solution \cite{Weiss.12.NULL}. 

The SBM consists of a two-level spin in contact with a bosonic heat bath and has the following Hamiltonian \cite{Leggett.87.RMP,Weiss.12.NULL}
\begin{equation}
\hat{H}=-\Delta \hat{\sigma}_x+\frac{1}{2}\sum_{j=1}^N\left(\hat{P}_j^2+\omega_j^2\hat{R}_j^2-2C_j\hat{R}_j\hat{\sigma}_z\right),
\end{equation}
where $\hat{\sigma}_{x/z}$ are the Pauli spin matrices, $\Delta$ is the tunnelling frequency between the two spin states, $\omega_j$ is the frequency of the $j$th harmonic oscillator, and $C_j$ is the coupling coefficient between the spin and the $j$th harmonic oscillator. To characterize the influence of the heat bath on the two-level subsystem, we employ an Ohmic spectral density with an exponential cutoff, namely $J(\omega)=\frac{\xi}{2}\pi\omega e^{-\omega/\omega_c}$, where the Kondo parameter $\xi$ characterizes the subsystem-bath coupling strength and $\omega_c$ is the cutoff frequency. The corresponding Weyl-ordered Hamiltonian in the mixed Wigner-Heisenberg representation is given by
\begin{equation}\label{eq:h_sb}
\hat{H}_{W}=-\Delta \hat{\sigma}_x+\frac{1}{2}\sum_{j=1}^N\left(P_j^2+\omega_j^2R_j^2-C_jR_j\hat{\sigma}_z-C_j\hat{\sigma}_zR_j\right).
\end{equation}

\subsection{Quantum speed limit time}
To evaluate the QSL time in the SBM, we set the initial spin state as the spin-up state, such that $\hat{\rho}_0=|+\rangle\langle +|$, where $|\pm\rangle$ are the eigenstates of $\hat{\sigma}_z$. 
In this case, the quantum fidelity in Eq.~(\ref{eq:ff_definition}) becomes
\begin{equation}\label{eq:ff_sb}
\mathcal{F}(\hat{\rho}_0,\hat{\rho}_{\tau})~=~\langle \mathcal{\hat{P}}_{++}(\tau)\rangle=\frac{1}{2}(1+B_z(\tau)),
\end{equation}
where $B_m(\tau)\equiv\langle \hat{\sigma}_m(\tau)\rangle$ ($m=x,y,z$). Using the fact that the reduced density matrix of a two-level system can be expressed as $\hat{\rho}_t=\frac{1}{2}(\mathcal{I}+\sum_{m=x,y,z}B_m(t)\hat{\sigma}_m)$, with $\mathcal{I}$ the $2\times2$ identity matrix, Eq.~(\ref{eq:r2}) may be rewritten as 
\begin{equation}\label{eq:r2_sb}
\mathrm{Tr}_S\left(\frac{d}{dt}\hat{\rho}_t\right)^2~=~\frac{1}{2}\sum_{m=x,y,z}\left(\frac{d}{dt}B_m(t)\right)^2.
\end{equation}
Thus, the QSL time is fully determined by the ensemble averages of the Pauli matrices and their corresponding time derivatives.

It is interesting to note that one may derive an analytical expression for $\tau_{QSL}$ in the isolated limit (i.e., $\hat{H}_I=0$) where the spin dynamics is governed by the Bloch equations \cite{Weiss.12.NULL}: $\frac{d}{dt}B_{x}(t)~=~0,~~\frac{d}{dt}B_{y}(t)~=~2\Delta B_z(t),~~\frac{d}{dt}B_{z}(t)~=~-2\Delta B_y(t)$.  Choosing the spin-up state as the initial state, the solution turns out to be $B_{x}(t)=0$, $B_{y}(t)=\sin(2\Delta t)$ and $B_{z}(t)=\cos(2\Delta t)$. Therefore, the QSL time for an isolated two-level subsystem is 
\begin{equation}\label{eq:tqslI}
\tau_{QSL}^{iso}~=~\frac{1}{2}\frac{1-\cos(2\Delta\tau)}{\sqrt{2}\Delta}.
\end{equation}
This will be used as a reference point to assess the effect of the subsystem-bath interaction on the QSL time.

\subsection{Computational details} 
We now describe the details of our DECIDE simulations.  In addition, to pinpoint the effects of stronger subsystem-bath coupling and non-Markovianity on the QSL times, we compare our DECIDE results (which capture these effects to a certain extent) to those of the perturbative Bloch-Redfield master equation with and without the Markov approximation.

\subsubsection{DECIDE simulations}
The generalized coordinates for the subsystem are taken to be the Pauli matrices, viz., $\boldsymbol{\hat{x}}=(\hat{\sigma}_x,\hat{\sigma}_y,\hat{\sigma}_z)$. The DECIDE EOMs for the subsystem and bath coordinates are derived from Eq.~(\ref{eq:eom_detail}) using $\hat{H}_W$ from Eq.~(\ref{eq:h_sb}) (the resulting EOMs are listed in Eq.~(\ref{eq:eom_sb}) in the Appendix). To compute the QSL time, we must calculate the ensemble averages $B_m(t)$ and their time derivatives (taking $\{|\alpha\rangle\}=\{|+\rangle,|-\rangle\}$ for convenience and the initial subsystem state to be $\hat{\rho}_0=|+\rangle\langle +|$) according to,
 \begin{eqnarray}\label{eq:btt}
B_{m}(t) &=& \int\,d\boldsymbol{X}(0)\rho_{B,W}(\boldsymbol{X}(0))\sigma_{m}^{++}(t),\nonumber\\
\frac{d}{dt}B_{m}(t) &=& \int\,d\boldsymbol{X}(0)\rho_{B,W}(\boldsymbol{X}(0))\frac{d}{dt}\sigma_{m}^{++}(t).
\end{eqnarray}
We evaluate the right-hand-sides of the above equations by averaging over a swarm of independent classical-like trajectories, with each trajectory starting from different values of the bath coordinates and the same values of the Pauli matrix elements.  More specifically, the initial values of the bath coordinates are $\boldsymbol{X}^{\alpha\alpha^{\prime}}(0)=\boldsymbol{X}(0)\delta_{\alpha\alpha^{\prime}}$ (due to the factorized initial state), with $\boldsymbol{X}(0)$ sampled from $\rho_{B,W}(0)$ in Eq.~(\ref{eq:rhoBw}). The Ohmic bath spectral density with an exponential cutoff is discretized as \cite{Thompson.99.JCP,Wang.01.JCP} 
\begin{equation}\label{eq:co}
C_j~=~\sqrt{\xi\hbar\omega_0}\omega_j,~~~\omega_j=-\omega_c\ln\left(1-j\frac{\omega_0}{\omega_c}\right),
\end{equation}
where $j$ runs from 1 to $N$, and $\omega_0=\frac{\omega_c}{N}(1-e^{-\omega_{max}/\omega_c})$ with $\omega_{max}$ the maximum frequency of the bath oscillators.  It should be noted that although we employ an Ohmic spectral density here, DECIDE, just like any other hybrid quantum-classical dynamics method, can handle arbitrary bath spectral densities.
Noting the forms of the Pauli matrices in the $\{|+\rangle,|-\rangle\}$ basis, it is easy to show that the non-vanishing initial values of the Pauli matrix elements are $\sigma_x^{-+}(0)=\sigma_x^{+-}(0)=1, \sigma_y^{-+}(0)=i, \sigma_y^{+-}(0) = -i, \sigma_z^{--}(0) =-1$, and $\sigma_z^{++}(0)=1$.  
Starting from the aforementioned initial conditions, we then numerically integrate Eq.~(\ref{eq:eom_sb}) using the fourth-order Runge-Kutta method \cite{Dormand.80.JCAM}, which results in trajectories of $\sigma_m^{++}(t)$ and their time derivatives.  Finally, averaging $\sigma_m^{++}(t)$ and their time derivatives over the ensemble of trajectories yields the required ensemble averages for constructing the QSL time, $\tau_{QSL}$.


\subsubsection{Bloch-Redfield master equation simulations}
To understand the impacts of strong subsystem-bath interactions and Markovian dynamics, we compare the results of our DECIDE simulations to those of the perturbative
 non-Markovian Bloch-Redfield equation (NM-BRE) and Markovian Bloch-Redfield equation (M-BRE).  Perturbative methods offer a simple means of simulating
 the reduced dynamics of two-level quantum systems \cite{Breuer.07.NULL}. By treating the subsystem-bath interaction, $\hat{H}_I$, as a perturbation to the lowest nontrivial order, one can obtain the NM-BRE within the Born approximation \cite{Aslangul.86.JP,Thoss.01.JCP,Boudjada.14.JPCA},
\begin{eqnarray}\label{eq:nm_bre}
\frac{d}{dt}B_{x}(t) &=& -\int_0^tds\Gamma_x(s)-\int_0^tds\Gamma_{xx}(s)B_x(t-s),\nonumber\\
\frac{d}{dt}B_{y}(t) &=& 2\Delta B_z(t)-\int_0^tds\Gamma_{yy}(s)B_y(t-s),\nonumber\\
\frac{d}{dt}B_{z}(t) &=& -2\Delta B_y(t).
\end{eqnarray}
The kernels of these integro-differential equations are $\Gamma_x(t)=-\sin(2\Delta t)M_2(t)$, $\Gamma_{xx}(t)= \cos(2\Delta t)M_1(t)$, and $\Gamma_{yy}(t)=M_1(t)$, with $M_1(t)$ and $M_2(t)$ satisfying the following relation,
\begin{eqnarray}
M_1(t)+iM_2(t) &=& \frac{4}{\pi}\int d\omega J(\omega)[\coth(\beta\omega/2)\cos\left(\omega t\right)\nonumber\\
&&+i\sin\left(\omega t\right)].
\end{eqnarray}

Next, introducing the Markov approximation, one arrives at the simpler M-BRE, which involves the following set of time-local equations \cite{Thoss.01.JCP},
\begin{eqnarray}\label{eq:m_bre}
\frac{d}{dt}B_{x}(t) &=& -G_x(t)-G_{xx}(t)B_x(t),\nonumber\\
\frac{d}{dt}B_{y}(t) &=& 2\Delta B_z(t)-G_{yy}(t)B_y(t)-G_{yz}(t)B_z(t),\nonumber\\
\frac{d}{dt}B_{z}(t) &=& -2\Delta B_y(t),
\end{eqnarray}
where the time-dependent coefficients are given by $G_x(t)=-\int_0^tdt^{\prime}\sin(2\Delta t^{\prime})M_2(t^{\prime})$, $G_{xx}(t)=G_{yy}(t)=\int_0^tdt^{\prime}\cos(2\Delta t^{\prime})M_1(t^{\prime})$, and $G_{yz}(t)=-\int_0^tdt^{\prime}\sin(2\Delta t^{\prime})M_1(t^{\prime})$. After integrating Eqs.~(\ref{eq:nm_bre}) and (\ref{eq:m_bre}) using the fourth-order Runge-Kutta scheme \cite{Dormand.80.JCAM}, we compute the QSL time from the $B_m(t)$'s and their time derivatives.

\subsection{Numerical results} 
Our goal is to investigate the interplay of subsystem-bath coupling strength and bath temperature on the behaviour of the QSL time. Noting the regime of validity of DECIDE, here we focus on cases with $\Delta<\omega_c$ and $T>\Delta$, while we vary the coupling strength $\xi$. 

\begin{figure}[tbh!]
  \centering
  \includegraphics[width=1.05\columnwidth]{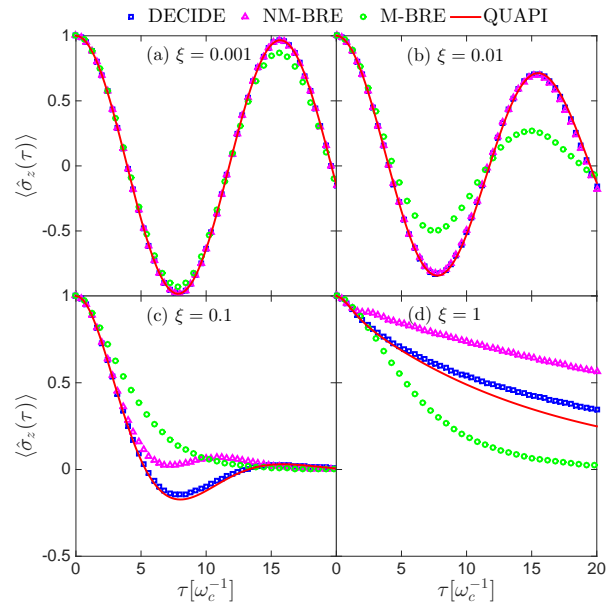}
\caption{Time-dependent spin polarization, $\langle \hat{\sigma}_z(\tau)\rangle$, for different subsystem-bath coupling strengths (a) $\xi=0.001$, (b) $\xi=0.01$, (c) $\xi=0.1$, and (d) $\xi=1$. Results obtained from QUAPI, DECIDE, M-BRE, and NM-BRE are represented by red solid lines, blue squares, green circles, and magenta triangles, respectively. To obtain converged results, an ensemble of $1\times10^4$ trajectories and MD time step of $\delta t=0.02$ were used in each DECIDE simulation. The values of the remaining parameters are $\Delta=0.2$, $T=1$, $\omega_c=1$, $\omega_{max}=5 \omega_c$, and $N=200$.}
\label{fig:sb_pop}
\end{figure}

\begin{figure}[tbh!]
  \centering
  \includegraphics[width=1\columnwidth]{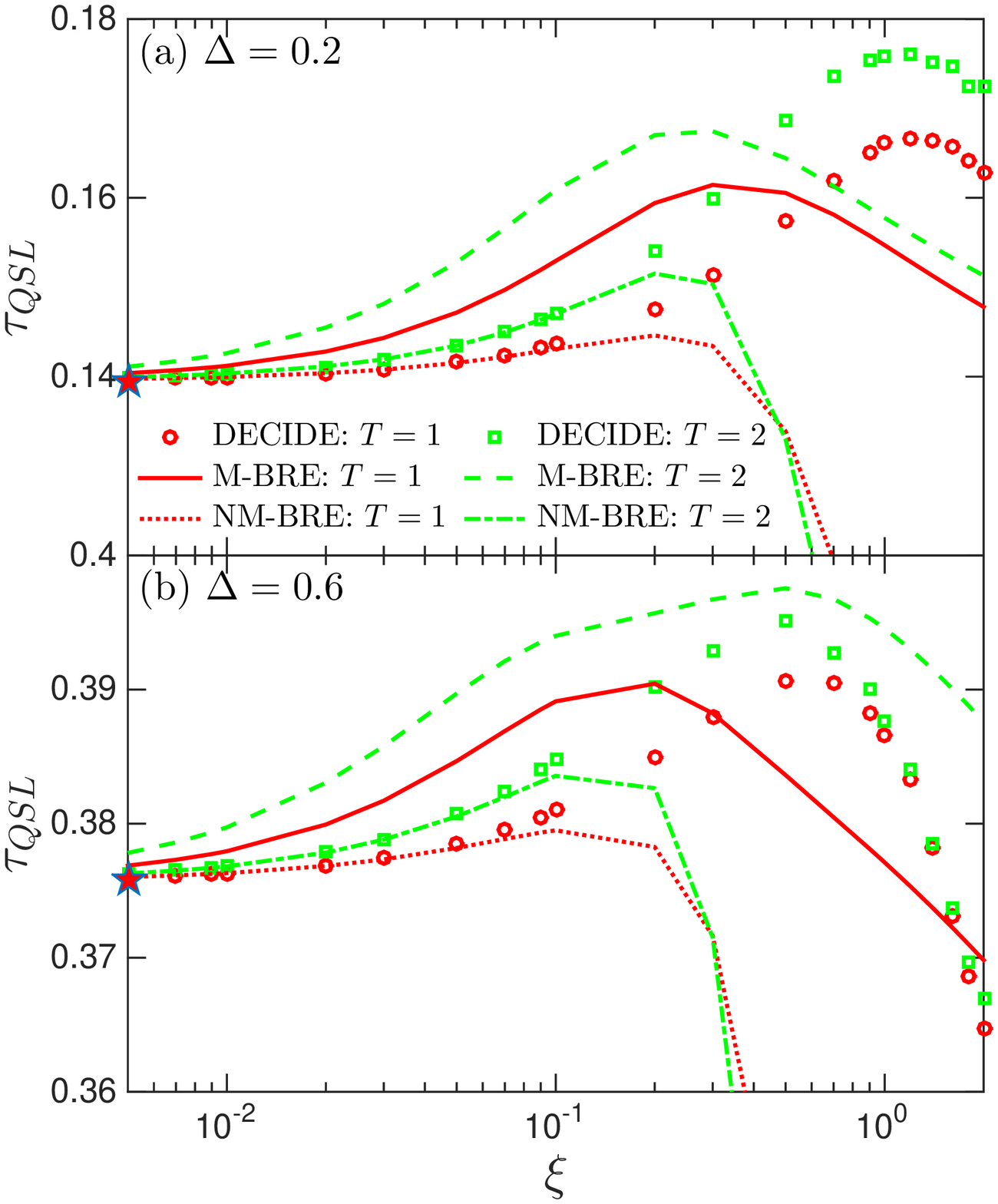}
\caption{QSL time, $\tau_{QSL}$, as a function of $\xi$ at different temperatures $T=1, 2$ for spin tunnelling frequencies (a) $\Delta=0.2$ and (b) $\Delta=0.6$, obtained using DECIDE (red circles and green squares), M-BRE (red solid lines and green dashed lines), and NM-BRE (red dotted lines and green dashed-dotted lines). 
The red stars on the y-axis correspond to the isolated QSL time, $\tau_{QSL}^{iso}$, from Eq.~(\ref{eq:tqslI}). To obtain converged DECIDE results, an ensemble of $1\times10^4$ trajectories and MD time step of $\delta t=0.005$ were used in each simulation.  In all simulations, the evolution time was fixed to be $\tau=1$. The values of the remaining parameters are $\omega_c=1$, $\omega_{max}=5 \omega_c$, and $N=200$.}
\label{fig:sb_qsl_d}
\end{figure}

To benchmark the performances of the various methods for simulating the reduced dynamics of the two-level system, we first compare in Fig.~\ref{fig:sb_pop} the spin polarization $\langle \hat{\sigma}_z(\tau)\rangle$ as obtained from NM-BRE, M-BRE, and DECIDE, with those calculated using the numerically exact QUAPI method \cite{Makri.95.JCP,Makri.95.JCPa}.
From the comparison, it is evident that M-BRE can only be applied in the very weak coupling regime, as even for $\xi=0.001$ we observe deviations from the numerically exact QUAPI results. By taking the non-Markovianity into account, NM-BRE improves upon M-BRE in the weak coupling regime as seen in panels (a)-(b) of Fig.~\ref{fig:sb_pop}. While both M-BRE and NM-BRE cannot be applied in the strong coupling regime given their perturbative nature, as seen in panels (c)-(d) of Fig.~\ref{fig:sb_pop}, it is important to note that they predict distinct dynamical behaviours in the strong coupling regime, viz., M-BRE fails to capture the bath-induced spin polarization, while NM-BRE overestimates the spin polarization (see panel (d) of Fig.~\ref{fig:sb_pop}). We will return to this observation when interpreting the behaviour of the QSL time in Fig.~\ref{fig:sb_qsl_d}. In contrast, DECIDE works well across all of the coupling regimes, demonstrating its reliability in simulating the reduced dynamics and hence the QSL time.

We now turn to the study of the QSL time. In Fig.~\ref{fig:sb_qsl_d}, we depict $\tau_{QSL}$ with a fixed evolution time of $\tau=1$ as a function of the subsystem-bath coupling strength $\xi$. In addition, we consider two different spin tunnelling frequencies $\Delta=0.2$ (Fig.~\ref{fig:sb_qsl_d} (a)) and $\Delta=0.6$ (Fig.~\ref{fig:sb_qsl_d} (b)).
Since the spin-boson system starts from a factorized initial state, NM-BRE and M-BRE can qualitatively capture the reduced dynamics at a very short time scale across the whole coupling regime as seen in Fig.~(\ref{fig:sb_pop}). Therefore, it is not surprising that we observe similar turnover behaviours of $\tau_{QSL}$ as a function of the coupling strength $\xi$ using all three methods. However, we should emphasize that only the predictions of DECIDE are reliable in the strong coupling regime, given the results in Fig.~\ref{fig:sb_pop}. When we compare the results of NM-BRE and M-BRE in the weak coupling regime, we find that the former predicts a smaller value of $\tau_{QSL}$ than the latter, for both values of the temperatures and spin tunnelling frequencies, which is a direct demonstration of non-Markovianity-induced speed-up \cite{Deffner.13.PRL}. We also note the perfect agreement between the DECIDE and NM-BRE results in the weak coupling regime, pointing to the importance of non-Markovianity even at high temperatures and weak subsystem-bath couplings, where it is often believed that non-Markovianity is minimal.  

When we increase the spin tunnelling frequency from $\Delta=0.2$ to $\Delta=0.6$, we see that DECIDE predicts a sharper decrease in the QSL time in the strong coupling regime (cf.~Figs.~\ref{fig:sb_qsl_d} (a) and (b)), and this leads to $\tau_{QSL}<\tau_{QSL}^{iso}$. This suggests that one could speed up the quantum evolution of an open quantum system by entering into the strong coupling regime with large tunnelling frequencies. This result is also a direct consequence of the non-Markovianity-induced speed-up \cite{Deffner.13.PRL}, since a larger spin tunnelling frequency corresponds to a larger ratio of $\Delta/\omega_c$ and consequently a stronger non-Markovian effect. As for the role of temperature in the QSL time, in Fig.~\ref{fig:sb_qsl_d} we observe that both M-BRE and DECIDE predict an increase in $\tau_{QSL}$ with increasing temperature for all couplings, while NM-BRE predicts an inverse temperature dependence in the very strong coupling regime. In the strong coupling regime, the spin is more polarized at the lower temperature, which implies a shorter QSL time $\tau_{QSL}$ at lower temperatures. Therefore, NM-BRE fails to capture the correct temperature dependence of $\tau_{QSL}$ in the strong coupling regime, while M-BRE does. This result demonstrates that since the perturbative and Markov approximations are linked, it is sometimes more consistent to enforce them together (as in the M-BRE) than separately.

\begin{figure}[tbh!]
  \centering
  \includegraphics[width=1\columnwidth]{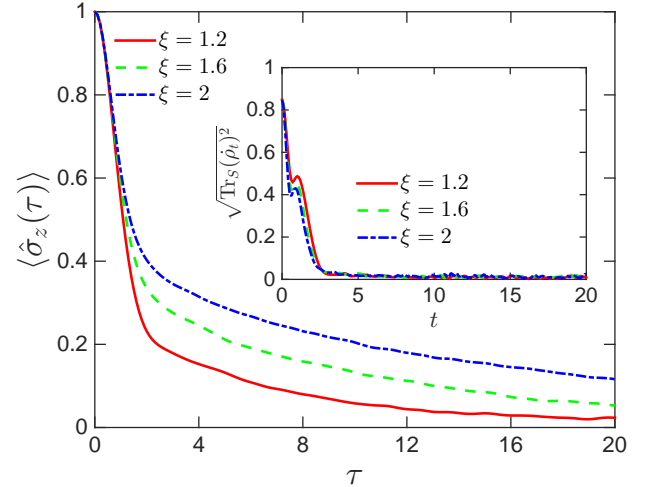}
\caption{Spin polarization, $\langle \hat{\sigma}_z(\tau)\rangle$, as a function of $\tau$ 
for different coupling strengths $\xi=1.2$ (red solid line), $1.6$ (green dashed line), and $2$ (blue dashed-dotted line). The inset depicts the time dependence of $\sqrt{\mathrm{Tr}_S(\dot{\rho}_t)^2}$ with the same coupling strengths. To obtain converged results, an ensemble of $1\times10^4$ trajectories and MD time step of $\delta t=0.01$ were used in each DECIDE simulation. The values of the remaining parameters are $\Delta=0.6$, $T=1$, $\omega_c=1$, $\omega_{max}=5 \omega_c$, and $N=200$.}
\label{fig:sb_strong}
\end{figure}

Qualitatively speaking, the turnover behaviour of the QSL time at higher coupling values, which is reminiscent of the turnover behaviour of the thermal conductance observed in the SBM \cite{Boudjada.14.JPCA,Liu.17.PRE}, results from the rapid energy exchange processes between the subsystem and bath in the strong coupling regime.  We can gain quantitative insight into the reduction of the QSL time by looking at the time dependences of the spin polarization, $\langle \hat{\sigma}_z(\tau)\rangle$, and $\sqrt{\mathrm{Tr}_S(\dot{\rho}_t)^2}$, as calculated by DECIDE, in the strong coupling regime (see Fig.~\ref{fig:sb_strong}).  We find that the stronger the coupling strength, the larger is the spin polarization. This monotonically increasing behaviour in the spin polarization implies a monotonically decreasing trend in the numerator of Eq.~(\ref{eq:tqsl}) (see also Eq.~(\ref{eq:ff_sb}) in the strong coupling regime. On the other hand, the integrand in the denominator of Eq.~(\ref{eq:tqsl}), $\sqrt{\mathrm{Tr}_S(\dot{\rho}_t)^2}$, is almost independent of coupling strength in the strong coupling regime.  
 
\begin{figure}[tbh!]
  \centering
  \includegraphics[width=1\columnwidth]{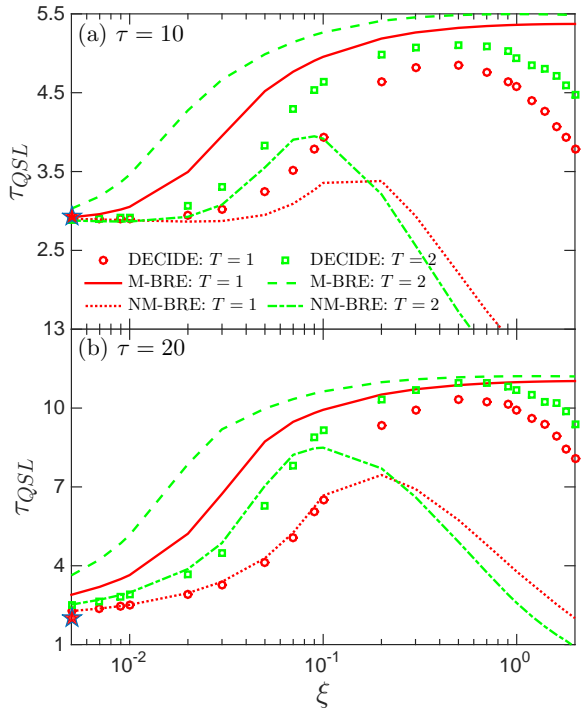}
\caption{QSL time, $\tau_{QSL}$, as a function of $\xi$ at different temperatures $T=1, 2$ for evolution times (a) $\tau=10$ and (b) $\tau=20$, obtained with DECIDE (red circles and green squares), M-BRE  (red solid lines and green dashed lines), and NM-BRE (red dotted lines and green dashed-dotted lines). 
The red stars on the y-axis correspond to the isolated QSL time, $\tau_{QSL}^{iso}$, from Eq.~(\ref{eq:tqslI}). To obtain converged DECIDE results, an ensemble of $1\times10^4$ trajectories and MD time step of $\delta t=0.005$ were used in each simulation.
The values of the remaining parameters are $\Delta=0.2$, $\omega_c=1$, $\omega_{max}=5 \omega_c$, and $N=200$.}
\label{fig:sb_qsl_d1}
\end{figure}

In Fig.~\ref{fig:sb_qsl_d1}, we present results of the QSL time at different temperatures as a function of the coupling strength, but now using longer evolution times $\tau$. We find that an increase in the evolution time from $\tau=1$ to $\tau=10$ and $\tau=20$, does not induce qualitative changes in the predictions of DECIDE compared to what we found in Fig.~\ref{fig:sb_qsl_d}.  
However, $\tau_{QSL}$ obtained from M-BRE now exhibits a monotonically increasing trend, as M-BRE fails to accurately describe the reduced dynamics at long times in the strong coupling regime. Although NM-BRE can still predict a turnover behaviour, the temperature dependence of $\tau_{QSL}$ in the strong coupling regime, as determined by NM-BRE, is inaccurate.

\section{The Fenna-Matthews-Olson complex}\label{sec:3}

There has been much interest in studying the dynamics of excitation energy transfer (EET)
in light-harvesting complexes, with a focus on the question of whether or not electronic quantum coherences assist the highly efficient transfer of energy in the photosynthetic process.  In particular, the FMO complex has gained much attention, with early 2D electronic spectroscopy experiments suggesting the existence of long-lived coherences at 77 K \cite{Engel.07.Nature}, and more recent studies disagreeing with the interpretation of the early experimental results \cite{Miller.17.pnas}. Apart from the debate over the existence of these long-lived coherences and their potential impact on the energy transfer efficiency, there has been a multitude of theoretical and computational studies on the FMO complex model aimed at benchmarking new methodologies and questioning the role of coherences in quantum dynamics \cite{Ishizaki.09.PNAS,Ishizaki.09.JCP,Nalbach.10.JCP,Ke.16.JCP}. Herein, we demonstrate the computation of QSL times for the FMO model and explore the relationship between these times and the EET efficiency at different temperatures.

\subsection{Model}
The photosynthetic FMO complex can be described by the following Frenkel exciton Hamiltonian in the single-excitation subspace \cite{Ishizaki.09.PNAS}
\begin{eqnarray}
\hat{H} &=& \sum_{n}\sum_{j=1}^M\left[\frac{\hat{P}_{n,j}^2}{2}+\frac{\omega_{n,j}^2}{2}\left(\hat{R}_{n,j}-\frac{C_{n,j}}{\omega_{n,j}^2}\mathcal{\hat{P}}_{nn}\right)^2\right]\nonumber\\
&&+\sum_{n}E_n\mathcal{\hat{P}}_{nn}+\sum_{m\neq n}V_{mn}\mathcal{\hat{P}}_{nm}.
\end{eqnarray}
In the above Hamiltonian, $\mathcal{\hat{P}}_{nm}=|n\rangle\langle m|$, where the basis state $|n\rangle$ corresponds to the $n$th chromophoric site in its electronically excited state and the remaining sites in their electronic ground state; $E_n$ denotes the excited state energy corresponding to state $|n\rangle$; and $V_{mn}$ is the excitonic coupling strength between the $n$th and $m$th sites. Each site is coupled to an independent heat bath containing $M$ harmonic oscillators at temperature $T$. We assume that the spectral densities of all the baths are equivalent, and can be characterized by Debye-Drude functions $J(\omega)=2\lambda_D\frac{\omega\tau_c}{1+\omega^2\tau_c^2}$ \cite{Ishizaki.09.PNAS}, where $\lambda_D$ is the bath reorganization energy and $\tau_c$ is the characteristic time. We adopt the values $\lambda_D=35~\mathrm{cm}^{-1}$ and $\tau_c=50~\mathrm{fs}$, as they yield excellent agreement between the experimental data and numerical simulations \cite{Read.08.BJ}.

As an illustration, we consider the apo-FMO complex, which includes seven bacteriochlorophyll (BChl) pigment-proteins per subunit; the conventional numbering of the BChls has been used here. The values of the site energies $E_n$ and excitonic coupling strengths $V_{mn}$ for a subunit of the apo-FMO complex \cite{Adolphs.06.BJ,Ishizaki.09.PNAS} are listed in Table \ref{tab:1}.
\begin{table}[tbh!]
\caption{Site energies $E_n$(diagonal entries) and excitonic coupling strengths $V_{mn}$ (off-diagonal entries) in units of $\mathrm{cm}^{-1}$ for a subunit of the apo-FMO complex}
\centering
\begin{tabular}{cccccccc}
\hline\hline
BChl & 1 & 2 & 3 & 4 & 5 & 6 & 7 \\ [0.5ex]
\hline
1 & 12410 & -87.7 & 5.5 & -5.9 & 6.7 & -13.7 & -9.9 \\
2 & -87.7 & 12530 & 30.8 & 8.2 & 0.7 & 11.8 & 4.3 \\
3 & 5.5 & 30.8 & 12210 & -53.5 & -2.2 & -9.6 & 6.0 \\
4 & -5.9 & 8.2 & -53.5 & 12320 & -70.7 & -17.0 & -63.3 \\
5 & 6.7 & 0.7 & -2.2 & -70.7 & 12480 & 81.1 & -1.3 \\
6 & -13.7 & 11.8 & -9.6 & -17.0 & 81.1 & 12630 & 39.7 \\
7 & -9.9 & 4.3 & 6.0 & -63.3 & -1.3 & 39.7 & 12440 \\ [1ex]
\hline\hline
\end{tabular}
\label{tab:1}
\end{table}
The corresponding Weyl-ordered Hamiltonian in the mixed Wigner-Heisenberg representation takes the form
\begin{eqnarray}\label{eq:h_fmo}
\hat{H}_W &=& \sum_{n,m=1}^7V_{nm}\mathcal{\hat{P}}_{nm}+\frac{1}{2}\sum_{n=1}^7\sum_{j=1}^M\left(P_{n,j}^2+\omega_{n,j}^2R_{n,j}^2\right)\nonumber\\
&&-\frac{1}{2}\sum_{n=1}^7\sum_{j=1}^MC_{n,j}\left(\mathcal{\hat{P}}_{nn}R_{n,j}+R_{n,j}\mathcal{\hat{P}}_{nn}\right),
\end{eqnarray}
where $V_{nn}= E_n+\sum_{j=1}^M\frac{C_{n,j}^2}{2\omega_{n,j}^2}$. For convenience and without loss of generality, we assume that the single excitation is initially located at site 1, such that $\hat{\rho}_0=|1\rangle\langle 1|$. As a result, the quantum fidelity in Eq.~(\ref{eq:ff_definition}) reduces to the simple form,
\begin{equation}\label{eq:ff_fmo}
\mathcal{F}(\hat{\rho}_0,\hat{\rho}_{\tau})~=~\langle \mathcal{\hat{P}}_{11}(\tau)\rangle.
\end{equation}
Equation (\ref{eq:r2}) remains the same with $L=7$. 

It was previously found that an initial excitation at BChl 1 rapidly transfers to BChls 3 and 4 (collectively known as the target region because they make contact with the reaction center complex) according to the following EET pathway \cite{Ishizaki.09.PNAS},
\begin{equation}
\mathrm{BChls}~1\to2\to3\rightleftharpoons4,
\end{equation}
where the double arrow implies that the excitation energy equilibrates between BChls 3 and 4 after BChl 3 is populated. 
\subsection{Simulation details}
It is well known that standard Redfield-type methods fail in describing EET even for dimers \cite{Ishizaki.09.JCP,Nalbach.10.JCP}. Therefore, we only use DECIDE to simulate the FMO model. The DECIDE EOMs for the subsystem and bath coordinates are listed in Eq.~(\ref{eq:eom_fmo}) in the Appendix.
The only non-zero element of the initial subsystem density matrix is $\rho_0^{11}=1$.  Based on Eq.~(\ref{eq:ea}) and taking $\{|\alpha\rangle\}=\{|1\rangle,|2\rangle, \ldots, |7\rangle\}$, the ensemble averages of the projection operators and their time derivatives, which are involved in the QSL time, are given by
\begin{eqnarray}\label{eq:p_fmo}
\langle \mathcal{\hat{P}}_{nm}(t)\rangle &=& \int\,d\boldsymbol{X}(0)\rho_{B,W}(\boldsymbol{X}(0))\mathcal{P}_{nm}^{11}(t),\nonumber\\
\frac{d}{dt}\langle \mathcal{\hat{P}}_{nm}(t)\rangle &=& \int\,d\boldsymbol{X}(0)\rho_{B,W}(\boldsymbol{X}(0))\frac{d}{dt}\mathcal{P}_{nm}^{11}(t),
\end{eqnarray}
To compute these ensemble averages, we generate a swarm of independent classical-like trajectories, with each trajectory starting from different values of the bath coordinates and the same values of the Pauli matrix elements.  More specifically, the initial values of the bath coordinates are $\boldsymbol{X}^{\alpha\alpha^{\prime}}(0)=\boldsymbol{X}(0)\delta_{\alpha\alpha^{\prime}}$ (due to the factorized initial state), with $\boldsymbol{X}(0)$ sampled from $\rho_{B,W}(0)$, which is now a product of seven partially Wigner-transformed Gaussian distributions (each given by Eq.~(\ref{eq:rhoBw})).  The Debye-Drude spectral density is discretized as \cite{Wang.01.JCP}
\begin{equation}
C_{n,j}~=~2\sqrt{\lambda_D\arctan(\omega_{max}\tau_c)/(\pi M)}\omega_{n,j},
\end{equation}
where $\omega_{n,j}=\tan\left[j\arctan(\omega_{max}\tau_c)/M\right]/\tau_c$. 
In the basis $\{|\alpha\rangle\}=\{|1\rangle,|2\rangle, \ldots, |7\rangle\}$, the initial nonzero values of the subsystem projection operator matrix elements, $\mathcal{P}_{nm}^{nm}(0)$, are 
\begin{equation}
\mathcal{P}_{nm}^{nm}(0)~=~\langle n|\hat{\mathcal{P}}_{nm}|m\rangle =1,~~\mathrm{for}~n, m \in\{1,2,\ldots,7\}.
 \end{equation}
Starting from the aforementioned initial conditions, we then numerically integrate Eq.~(\ref{eq:eom_fmo}) using the fourth-order Runge-Kutta scheme \cite{Dormand.80.JCAM}, which results in trajectories of $\mathcal{P}_{nm}^{11}(t)$ and their time derivatives.  Finally, averaging $\mathcal{P}_{nm}^{11}(t)$ and their time derivatives over the ensemble of trajectories yields the required ensemble averages for constructing the QSL time, $\tau_{QSL}$.

\subsection{Numerical results}
The validity of the DECIDE method for treating the FMO complex model was established in Ref.~\cite{Liu.18.JPCL}.  In Fig.~3 of Ref.~\cite{Liu.18.JPCL}, the time-dependent populations for the first four BChl pigment-proteins (the populations of the remaining pigments remain very small) obtained using DECIDE are compared with those obtained using the numerically exact forward-backward stochastic Schr\"odinger equation (FB-SSE) \cite{Ke.16.JCP}. As can be seen, DECIDE works well in describing the EET at both the physiological temperature (300 K) and the cryogenic temperature (77 K).  

\begin{figure}[tbh!]
  \centering
  \includegraphics[width=1\columnwidth]{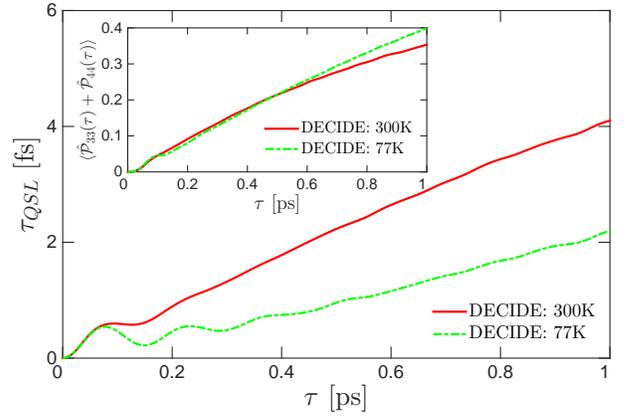}
\caption{QSL time, $\tau_{QSL}$, as a function of the evolution time $\tau$ at the physiological temperature of 300 K (red solid line) and cryogenic temperature of 77 K (green dashed-dotted line). The inset shows the combined population of the third and fourth BChl pigment-proteins (BChls 3 and 4), $\langle\hat{\mathcal{P}}_{33}(\tau)+\hat{\mathcal{P}}_{44}(\tau)\rangle$, at the physiological temperature (red solid line) and cryogenic temperature (green dashed-dotted line). An ensemble of $1\times10^4$ trajectories and MD time step $\delta t=$1 fs were used to obtain the converged DECIDE results. The values of the bath parameters are $M=40$, $\tau_c=50$ fs, and $\lambda_D=35~\mathrm{cm}^{-1}$.}
\label{fig:fmo}
\end{figure}

As the parameters of the FMO complex model were determined by numerically fitting the experimental data, we only vary the evolution time $\tau$ to study the QSL time, unlike in our calculations for the SBM.
We recall that the QSL time can be analyzed either by keeping $\tau$ fixed, or by studying the behaviour of the QSL time as a function of $\tau$.  In Fig.~\ref{fig:fmo}, we present
results for $\tau_{QSL}$ as a function of $\tau$ for both a physiological temperature (300 K) and a cryogenic temperature (77 K). 
At $\tau=0$, $\tau_{QSL}=0$ due to a unit quantum fidelity. As time progresses, we first observe oscillations in $\tau_{QSL}$ at both temperatures, with the oscillations at 77 K being more pronounced. At later times, after the oscillations have decayed, $\tau_{QSL}$ is seen to grow almost linearly with $\tau$. Apart from the very short time regime, the FMO complex at 77 K has a shorter QSL time, or equivalently, a faster quantum evolution speed. Since the QSL time sets the bound on the minimal evolution time for an initial quantum state to reach a target state, a shorter QSL time should correspond to a faster EET process between BChl 1 and the target region.  The fact that $\tau_{QSL} \ll \tau$ here is noteworthy and warrants further investigation.

To understand the significance of a faster quantum evolution speed at 77 K, we recall that BChls 3 and 4 define the target region in contact with the reaction center complex. The total population accumulating in time at sites 3 and 4 at both 77 K and 300 K is shown in the inset of Fig.~\ref{fig:fmo}. As can be seen from the inset, the results for the two temperatures are very similar up to $\sim 0.6$ ps, after which the total population at 77 K begins to exceed that at 300 K.  By $\tau=1$ ps, the total population at 77 K is slightly larger than that at 300 K (0.4 vs.~0.35, respectively).  On the other hand, the QSL times at these two temperatures begin to show deviations at a very short time and differ by about a factor of two at $\tau=1$ ps.  These differences are a reflection of the fact that the populations only convey a portion of the dynamical information contained in the QSL time (as the QSL time also contains information about the coherences).  Now, since only the excitation energy that has been transferred into the target region can be utilized in the reaction center, the total population accumulation at BChls 3 and 4 essentially determines the efficiency of the EET process.  Therefore, by comparing our QSL times to the total populations of BChls 3 and 4 at the two different temperatures, our results suggest that a faster quantum evolution speed could lead to a more efficient EET in the long-time limit.  For this reason, studying the behaviours of QSL times may allow one to discover useful design principles for creating efficient artificial light-harvesting devices.

\section{Summary}\label{sec:4}

Hybrid quantum-classical dynamics methods have been primarily developed to simulate quantum processes in condensed phase systems, such as chemical reaction dynamics and vibrational energy relaxation. 
The main goal of our paper was to offer a hybrid quantum-classical method as a powerful tool for studying performance bounds in open quantum systems. 
Specifically, we simulated the quantum speed limit in multi-level open quantum systems, using a hybrid quantum-classical method that is based on the mixed Wigner-Heisenberg representation of the composite quantum dynamics.  Within this approach, the information over the QSL time is encoded into a set of coupled first-order differential equations for the quantum and classical degrees of freedom. The flexibility inherent to this method allowed us to explore systems that go beyond very simple models and parameters regimes that could not be examined with low-order perturbative quantum master equations. 

We studied QSL times in two models. First, we considered the spin-boson model, a prototype open quantum system. The effects of bath temperature and subsystem-bath coupling strength on the behaviour of the QSL time were analyzed. In particular, we found that the QSL time exhibits a turnover behaviour as a function of the subsystem-bath coupling strength, which we attributed to the strong dissipation effect in the strong coupling regime. Under certain circumstances, this turnover could even lead to a speed-up of the quantum evolution in the strong coupling regime as compared to in the isolated limit. By comparing the results from the hybrid quantum-classical method with those obtained from NM-BRE and M-BRE, we concluded that perturbative methods should not be used to calculate QSL times beyond their strict regimes of validity.

As a second example, we studied the QSL time in the Fenna-Matthews-Olson complex. By varying the evolution time, we found that the quantum evolution speed increases as we reduce the temperature. We suggested that this faster quantum evolution speed could lead to a faster excitation energy transfer (EET) to the reaction center complex.

Future studies will aim at gaining a better understanding of the role of the QSL time in open quantum systems using the hybrid quantum-classical method. We also anticipate the development of potential quantum control protocols for EET applications by identifying key characteristics of the QSL time in a given EET process.

\begin{acknowledgments}
J.~Liu and G.~Hanna acknowledge support from the Natural Sciences and Engineering Research Council of Canada (NSERC). D.~Segal acknowledges support from an NSERC Discovery Grant and the Canada Research Chair program.
\end{acknowledgments}

\renewcommand{\theequation}{A\arabic{equation}}
\setcounter{equation}{0}  
\appendix 
\section*{Appendix: DECIDE equations of motion}

The general EOMs in the DECIDE hybrid mixed quantum-classical method are given in Eq.~(\ref{eq:eom_detail}).  Here, we provide the specific DECIDE equations for the spin-boson and FMO complex models.

\subsection{Equations for the spin-boson model}\label{b:11}
From Eq.~(\ref{eq:eom_detail}), one can write down the EOMs for the quantum and classical coordinates using the form of the spin-boson Hamiltonian in Eq.~(\ref{eq:h_sb}).  This results in the following set of coupled FODEs for the matrix elements of the spin Pauli matrices and bath oscillator coordinates (taking $\{|\alpha\rangle\}=\{|+\rangle,|-\rangle\}$),
\begin{eqnarray}\label{eq:eom_sb}
\frac{d}{dt}\sigma_x^{\alpha\alpha^{\prime}}(t) &=& \sum_{j=1}^NC_{j}[R_{j}(t)\hat{\sigma}_y(t)+\hat{\sigma}_y(t)R_{j}(t)]^{\alpha\alpha^{\prime}},\nonumber\\
\frac{d}{dt}\sigma_y^{\alpha\alpha^{\prime}}(t) &=& 2\Delta\sigma_z^{\alpha\alpha^{\prime}}(t)-\sum_{j=1}^NC_{j}[R_{j}(t)\hat{\sigma}_x(t)\nonumber\\
&&+\hat{\sigma}_x(t)R_{j}(t)]^{\alpha\alpha^{\prime}},\nonumber\\
\frac{d}{dt}\sigma_z^{\alpha\alpha^{\prime}}(t) &=& -2\Delta\sigma_y^{\alpha\alpha^{\prime}}(t),\nonumber\\
\frac{d}{dt}R_{j}^{\alpha\alpha^{\prime}}(t) &=& P_{j}^{\alpha\alpha^{\prime}}(t),\nonumber\\
\frac{d}{dt}P_{j}^{\alpha\alpha^{\prime}}(t) &=& -\omega_{j}^2R_{j}^{\alpha\alpha^{\prime}}(t)+C_{j}\sigma_z^{\alpha\alpha^{\prime}}(t),
\end{eqnarray}
where the terms of the form $[R_{j}(t)\hat{\sigma}_m(t)]^{\alpha\alpha^{\prime}}$ are evaluated as $\sum_{\beta}R_{j}^{\alpha\beta}(t)\sigma_m^{\beta\alpha^{\prime}}(t)$. As can be seen, the above set of EOMs consists of $4\times(3+2N)$ coupled FODEs for the matrix elements $(\sigma_x^{\{\alpha\alpha^{\prime}\}},\sigma_y^{\{\alpha\alpha^{\prime}\}},\sigma_z^{\{\alpha\alpha^{\prime}\}},\boldsymbol{X}^{\{\alpha\alpha^{\prime}\}})$, where $N$ is the number of bath oscillators.

\subsection{Equations for the FMO complex model}\label{b:12}
From Eq.~(\ref{eq:eom_detail}), one can write down the EOMs for the quantum and classical coordinates using the form of the FMO complex Hamiltonian in Eq.~(\ref{eq:h_fmo}).  This results in the following set of coupled FODEs for the matrix elements of the chromophoric site projectors $\{\hat{\mathcal{P}}_{nm}\}$ (with $m,n\in\{1,2,\ldots,7\}$) and bath oscillator coordinates (taking $\{|\alpha\rangle\}=\{|1\rangle,|2\rangle, \cdots, |7\rangle\}$), 
\begin{eqnarray} \label{eq:eom_fmo}
&&\frac{d}{dt}\mathcal{P}_{nm}^{\alpha\alpha^{\prime}}(t)~=~i[\sum_{l=1}^7V_{ln}\mathcal{P}_{lm}^{\alpha\alpha^{\prime}}(t)-\sum_{k=1}^7V_{mk}\mathcal{P}_{nk}^{\alpha\alpha^{\prime}}(t)]\nonumber\\
&&-\frac{i}{2}\sum_{j=1}^MC_{n,j}\left[R_{n,j}(t)\hat{\mathcal{P}}_{nm}(t)+\hat{\mathcal{P}}_{nm}(t)R_{n,j}(t)\right]^{\alpha\alpha^{\prime}}\nonumber\\
&&+\frac{i}{2}\sum_{j=1}^MC_{m,j}\left[R_{m,j}(t)\hat{\mathcal{P}}_{nm}(t)+\hat{\mathcal{P}}_{nm}(t)R_{m,j}(t)\right]^{\alpha\alpha^{\prime}},\nonumber\\
&&\frac{d}{dt}R_{n,j}^{\alpha\alpha^{\prime}}(t) ~=~ P_{n,j}^{\alpha\alpha^{\prime}}(t),\nonumber\\
&&\frac{d}{dt}P_{n,j}^{\alpha\alpha^{\prime}}(t) ~=~ -\omega_{n,j}^2R_{n,j}^{\alpha\alpha^{\prime}}(t)+C_{n,j}\mathcal{P}_{nn}^{\alpha\alpha^{\prime}}(t),
\end{eqnarray}
where the terms of the form $[R_{n,j}(t)\hat{\mathcal{P}}_{nm}(t)]^{\alpha\alpha^{\prime}}$ are evaluated as $\sum_{\beta}R_{n,j}^{\alpha\beta}(t)\mathcal{P}_{nm}^{\beta\alpha^{\prime}}(t)$. 
Given the completeness condition $\sum_{n=1}^7\hat{\mathcal{P}}_{nn}=1$, there are $49\times(48+14M)$ coupled FODEs for the subsystem and bath matrix elements, where $M$ is the number of bath oscillators coupled to each site.
\bibliography{qsl}
\end{document}